\documentclass[10pt,conference]{IEEEtran}
\usepackage{amssymb}
\usepackage{amsmath}
\usepackage{cite}
\usepackage{url}
\usepackage{xcolor}
\usepackage{cite,graphicx,amsmath,amssymb}
\usepackage{subfigure}
\usepackage{cite}
\usepackage{caption}
\usepackage{amsthm}
\usepackage{algorithm}
\usepackage{algorithmic}
\allowdisplaybreaks

\usepackage[
top    = 0.75 in,
bottom = 1.05 in,
left   = 0.63 in,
right  = 0.63 in]{geometry}

\newtheorem{theorem}{Theorem}

\newtheorem{lemma}{Lemma}

\newtheorem{corollary}{Corollary}

\newcommand{\bm}[1]{\mbox{\boldmath{$#1$}}}

\usepackage{makecell}
\IEEEoverridecommandlockouts

\begin{document}

	\title{Relay-Assisted Activation-Integrated SIM for Wireless Physical Neural Networks}
	\author{
			\IEEEauthorblockN{Meng Hua  and   Deniz~G\"und\"uz}
		\IEEEauthorblockA{Department of Electrical and Electronic Engineering, Imperial College London, United Kingdom}
		Email: {{\{m.hua, d.gunduz\}}@imperial.ac.uk}	
		\thanks{This work was  supported by the SNS JU Project 6G-GOALS under the	EU’s Horizon Program with Grant  101139232.
		}
	}

\maketitle
\begin{abstract}
Wireless physical neural networks (WPNNs) have emerged as a promising paradigm for performing neural computation directly in the physical layer of wireless systems, offering  low latency and high energy efficiency. However, most existing WPNN implementations primarily rely on linear physical transformations, which fundamentally limits their expressiveness.
In this work, we propose a relay-assisted WPNN architecture based on \textit{activation-integrated stacked intelligent metasurfaces} (AI-SIMs), where each passive metasurface layer enabling linear wave manipulation is cascaded with an activation metasurface layer that realizes nonlinear processing in the analog domain. By deliberately structuring multi-hop wireless propagation, the relay amplification matrix and the metasurface phase-shift matrices jointly act as trainable network weights, while hardware-implemented activation functions provide essential nonlinearity.
Simulation results demonstrate that the proposed architecture achieves high classification accuracy, and that incorporating hardware-based activation functions significantly improves representational capability and performance compared with purely linear physical implementations.
\end{abstract}

\section{Introduction}
Wireless physical neural networks (WPNNs) have recently emerged as a promising paradigm that exploits native wireless propagation and hardware characteristics to perform inference directly in the physical layer \cite{hua2026WPNN,wright2022deep, iten2020discovering }. In  conventional digital neural network pipelines,  inference is executed in the digital domain on finite-precision data tensors, while wireless transmission is abstracted as a separate bit pipe that delivers representations. As a result, end-to-end systems typically rely on analog-to-digital and digital-to-analog conversions at the transceiver, introducing non-negligible additional latency and energy consumption. By contrast, WPNNs leverage the wireless channel together with reconfigurable hardware components as trainable physical transformations, thereby enabling inference to be carried out directly over the propagation medium through over-the-air (OTA) processing.

Among various candidates, two of the most promising trainable physical hardware platforms have attracted significant attention, namely metasurfaces \cite{hua2025aircnn,stylianopoulos2026over,stylianopoulos2025over,Garcia2023irNN} and relays \cite{girici2026realization,bergel2024nonlinear,bian2025overtheair,binyamini2025estimating,wang2022distributed}. Metasurfaces consist of a large number of subwavelength meta-atoms whose electromagnetic responses can be dynamically tuned, enabling programmable manipulation of wireless wavefronts, such as phase, amplitude, and polarization. For  example, work \cite{hua2025aircnn} adopted reconfigurable intelligent surfaces (RISs) and optimized phase shifts to engineer the ambient wireless propagation environment so as to emulate the operations of a convolutional neural network layer.
On the other hand, multiple distributed relays actively process and forward received signals by optimizing  amplification gains, behaving as  a fully-connected (FC) layer. For example, \cite{bergel2024nonlinear} treated the relay network itself, comprising  a large number of distributed relays, as a neural network to implement communication task by training relay gains. 
However, most existing WPNN implementations are dominated by \emph{linear} physical transformations, e.g., phase-only wave manipulation by RISs. Such linearity fundamentally limits the expressive capability of physical networks. In addition, metasurface-based implementations are inherently constrained by the unit-modulus phase-shift requirement, which further limits their neural expressiveness.

To address the above limitations, this paper proposes a relay-assisted WPNN architecture enabled by activation-integrated stacked intelligent metasurfaces (AI-SIMs). The proposed architecture integrates linear wave manipulation and hardware-implemented nonlinear activation within a unified OTA computational framework. Specifically, each AI-SIM layer consists of a linear metasurface layer with independently programmable phase shifts, followed by an activation metasurface layer that realizes analog-domain nonlinear processing via radio-frequency detection and analog circuits. This cascaded design enables energy-efficient linear transformations together with intrinsic hardware nonlinearity, thereby overcoming the expressiveness limitations of purely linear physical implementations. In addition, the relay is elevated from a communication facilitator to an integral computational layer of the WPNN. By explicitly accounting for hardware-imperfect relay power amplifiers, whose nonlinear input–output characteristics naturally resemble neural activation functions, the relay provides flexible gain control and effectively emulates FC layers.
Through structured OTA propagation, the metasurface phase-shift matrices and the relay amplification matrix jointly serve as trainable network weights, while hardware-implemented nonlinearities supply essential expressive power. Numerical results demonstrate that the proposed architecture achieves substantial accuracy gains over linear physical implementations, particularly in low signal-to-noise ratio (SNR) regimes.

\begin{figure}[!t]
	\centerline{\includegraphics[width=3.5in]{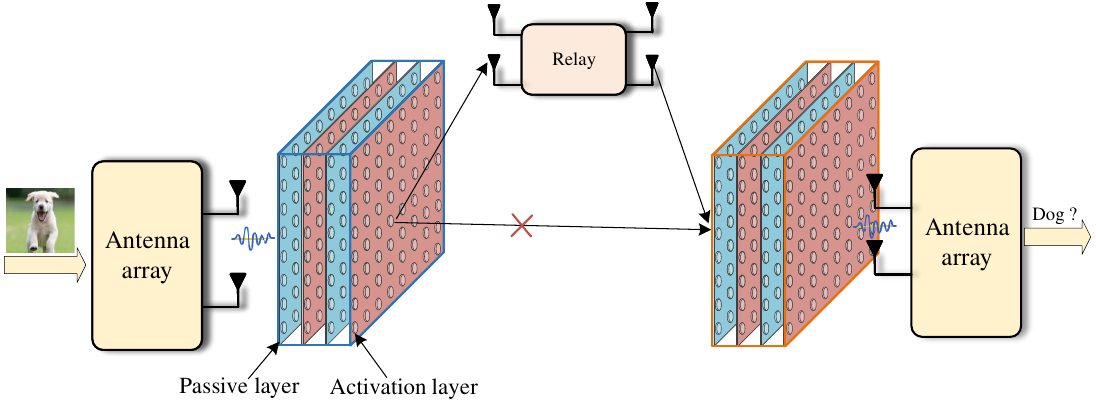}}
	\caption{Relay-assisted AI-SIM architecture for implementing WPNN.} \label{model_fig1}
\end{figure}

\section{System Model}
As illustrated in Fig.~\ref{model_fig1}, we consider a relay-assisted WPNN system in which both the transmitter and the receiver are equipped with AI-SIMs. Each AI-SIM comprises $L$ passive metasurface layers for linear electromagnetic wave manipulation, where each passive layer is immediately followed by an activation metasurface layer that implements nonlinear signal processing in the analog domain.

At the transmitter side, the information-bearing signals are radiated through an antenna array and subsequently processed by the AI-SIM. Similarly, at the receiver side, an AI-SIM is placed before the antenna array to process the impinging electromagnetic waves.
The direct wireless link between the transmitter and receiver is assumed to be blocked, resulting in a non-line-of-sight (NLoS) propagation condition. To enable reliable communication, a relay node is deployed to establish a multi-hop transmission link from the transmitter to the receiver.   The relay forwards the received signals between the transmitter- and receiver-side AI-SIMs, and can  be treated as  a physical FC layer. 

The WPNN is employed for classification tasks. Specifically, an input image, denoted by ${\bf S} \in {{\mathbb R}^{C \times H \times W}}$, where $C$, $H$, and $W$ represent the number of color channels, height, and width, respectively, is transmitted via a single relay. Let $N_{\rm t}$, $N_{\rm s}$, and $N_{\rm r}$ denote the numbers of transmit antennas, relay antennas, and receive antennas, respectively. For notational simplicity, we assume that the number of meta-atoms on each passive and activation metasurface is identical and equal to $M$. 

\subsection{Passive and Activation-based Metasurface}
Let ${\mathbf{\Theta }}_l^{\text{t}} = {\text{diag}}\left( {{e^{j\theta _{l,1}^{\text{t}}}}, \ldots ,{e^{j\theta _{l,M}^{\text{t}}}}} \right) \in {{\mathbb C}^{M \times M}}$, $l = \left\{ {1, \ldots ,L} \right\}$, denote the phase shift matrix of the $l$-th passive metasurface at the transmitter-side  AI-SIM, where $\theta _{l,m}^{\text{t}} \in \left[ {0,2\pi } \right)$ is the phase shift of  its $m$-th meta-atom.   Similarly, denote by ${\mathbf{\Theta }}_l^{\text{r}} = {\text{diag}}\left( {{e^{j\theta _{l,1}^{\text{r}}}}, \ldots ,{e^{j\theta _{l,M}^{\text{r}}}}} \right) \in {{\mathbb C}^{M \times M}}$, $l = \left\{ {1, \ldots ,L} \right\}$, the phase shift matrix of the $l$-th passive metasurface at the receiver-side  AI-SIM.
Each passive metasurface layer is followed by an activation-based metasurface layer with $M$ nonlinear meta-atoms. According to \cite{ning2025multilayer}, the activation-based metasurface layer can  integrate  radio-frequency detectors, amplifiers, and voltage adders, such that each nonlinear meta-atom can implement tanh-like behavior depending on the circuit configuration. To capture the resulting nonlinearity, we define  $\bar \pi \left(  \cdot  \right)$ as the nonlinear operator associated with the activation metasurface layer.

\subsection{Channel Model} 
We assume that each metasurface is arranged as a uniform planar array (UPA), while the transmit and receive antennas are configured as  uniform linear arrays (ULAs). Note that the metasurfaces  are closely stacked, such that  inter-layer propagation occurs in the near-field regime. According to  Rayleigh-Sommerfeld diffraction theory, the transmission coefficient from the $n$-th meta-atom on the $\left( {l - 1} \right)$-th metasurface layer   to the $m$-th meta-atom on the $l$-th metasurface layer at the transmitter-side  AI-SIM, $l = \left\{ {2, 3,\ldots ,2L} \right\}$,  denoted by ${\left[ {{{\mathbf{W}}_{l}^{\rm t}}} \right]_{m,n}}$,  can be  characterized as 
\begin{align}
{\left[ {{{\mathbf{W}}_l^{\rm t}}} \right]_{m,n}} = \frac{{{A}\cos \phi _{m,n}^{(l)}}}{{d_{m,n}^{(l)}}}\left( {\frac{1}{{2\pi d_{m,n}^{(l)}}} - \frac{j}{{{\lambda}}}} \right)\exp \left( {j\frac{{2\pi d_{m,n}^{(l)}}}{{{\lambda}}}} \right),
\end{align}
where ${d_{m,n}^{(1)}}$ denotes the transmission distance, $\lambda $ represents the wavelength, $A$ is the area of each meta-atom, and ${\phi _{m,n}^{(l)}}$ 
represents the angle between the propagation direction and
the normal direction of the $\left( {l - 1} \right)$-th metasurface layer.  Denote by ${\mathbf{W}_1^{\rm t}} \in {{\mathbb C}^{M \times {N_{\text{t}}}}}$ the  transmission coefficient matrix between the $N_{\rm t}$ transmit antennas  and $M$ meta-atoms in the first layer of the transmitter-side AI-SIM. 
Let ${\mathbf{X}}_0^{\text{t}} = \bf{S}_{\rm c}$ denote the input to the transmitter side AI-SIM. The output of the $l$-th activation layer is given by
 \begin{align}
     {\mathbf{X}}_l^{\text{t}} = \bar \pi \left( {\mathbf{W}}_{2l}^{\text{t}}{{\mathbf{\Theta }}_{2l-1}^{\text{t}}{\mathbf{W}}_{2l-1}^{\text{t}} {\mathbf{X}}_{l-1}^{\text{t}} \!+\! {\mathbf{b}}_{l}^{\text{t}}} {\mathbf{1}}^T \right),  l = 1, \ldots, L, \label{channel: transmitter}
 \end{align}
 where ${{\mathbf{b}}_l^{\text{t}}}\in {{\mathbb C}^{M \times 1}}$ is a trainable bias vector at the $l$-th activation metasurface layer realized by injecting  direct-current offsets \cite{ning2025multilayer}. 

 Similarly,  at the receiver-side  AI-SIM,  the transmission coefficient from the $n$-th meta-atom on the $\left( {l - 1} \right)$-th metasurface layer to the $m$-th meta-atom on the $l$-th metasurface layer is denoted by ${{{\mathbf{W}}_l^{\rm r}}}$, $l = \left\{ {2,3, \ldots ,2L} \right\}$. Let ${\mathbf{X}}_0^{\text{r}}$ denote the input to the receiver side AI-SIM. The cascaded processing is given by
 \begin{align}
     {\mathbf{X}}_l^{\text{r}} = \pi \left( {\mathbf{W}}_{2l}^{\text{r}}{{\mathbf{\Theta }}_{2l-1}^{\text{r}}{\mathbf{W}}_{2l-1}^{\text{r}} {\mathbf{X}}_{l-1}^{\text{r}} \!+\! {\mathbf{b}}_{l}^{\text{r}}} {\mathbf{1}}^T \right),  l = 1, \ldots, L, \label{channel: receiver}
 \end{align}
 where $\left\{ {\mathbf{b}}_l^{\text{r}} \right\}_{l=1}^{L}$ are trainable bias vectors and ${{{\mathbf{W}}_1^{\rm r}}}$ denotes the transmission coefficient matrix from the relay to the first layer of the receiver-side AI-SIM.


We denote by ${{\mathbf{G}}_{\text{t}}}$ and ${{\mathbf{G}}_{\text{r}}}$ the channel matrices from  the transmitter-side AI-SIM to the relay  and  from  the relay to the  receiver-side AI-SIM, respectively. For the relay-related links, the propagation distance is assumed to be large compared with the aperture size, such that signal transmission occurs in the far-field regime.
 Thus, we consider a Rayleigh fading model for  the relay-related channels, where each entry of ${{\mathbf{G}}_{\text{s}}}$ is independently distributed as  ${\left[ {{{\mathbf{G}}_{\rm s}}} \right]_{i,j}} \sim {\cal CN}\left( {0,1 } \right), {\rm s}\in \{{\rm t},{\rm r}\}$.

\subsection{WPNN-based End-to-End Signal Representation}
The image $\bf S$ is first mapped into a complex-valued matrix, denoted by
${{\bf{S}}_{\rm c}} \in {{\mathbb C}^{{N_{ \rm t}} \times \frac{{CHW}}{{2{N_{\text{t}}}}}}}$, which serves as the input to the transmitter-side AI-SIM.

The transmitter performs an encoding operation, denoted by
\begin{align} 
{\mathcal E}\left(  \cdot  \right):{{\mathbb C}^{{N_{\text{t}}} \times \frac{{CHW}}{{2{N_{\text{t}}}}}}} \to {{\mathbb C }^{M \times \frac{{CHW}}{{2{N_{\text{t}}}}}}},
\end{align}
which  maps ${{\bf{S}}_{\rm c}}$ directly into latent features ${{\mathbf{S}}_{\text{t}}} \in {{\mathbb C}^{{M} \times \frac{{CHW}}{{2{N_{\text{t}}}}}}}$.  

At the relay node, the received features are processed through an intermediate physical-layer mapping, denoted by
\begin{align} 
{\cal R}\left(  \cdot  \right):{{\mathbb C}^{M \times \frac{{CHW}}{{2{N_{\text{t}}}}}}} \to {{\mathbb C}^{{N_{\text{s}}} \times \frac{{CHW}}{{2{N_{\text{t}}}}}}},
\end{align}
which transforms ${{\bf{S}}_{\rm t}}$ into relay-domain features ${{\mathbf{S}}_{\text{s}}} \in {{\mathbb C}^{N_{\rm s} \times \frac{{CHW}}{{2{N_{\text{t}}}}}}}$.

At the receiver side, a decoding operation is applied to the received features, denoted by
\begin{align} 
{\cal D}\left(  \cdot  \right):{{\mathbb C}^{N_{\rm s} \times \frac{{CHW}}{{2{N_{\text{t}}}}}}} \to {{\mathbb C}^{{N_{\text{r}}} \times \frac{{CHW}}{{2{N_{\text{t}}}}}}},
\end{align}
which produces the task-oriented feature representation  ${{\mathbf{S}}_{\text{r}}} \in {{\mathbb C}^{{N}_{\rm r} \times \frac{{CHW}}{{2{N_{\text{t}}}}}}}$ for subsequent classification.

The encoder, the intermediate mapper, and  the decoder are parameterized by WPNNs, yielding
\begin{align} 
{{\mathbf{S}}_{\text{t}}} = {{\cal E}_{{{\bm{\varphi }}_{\text{t}}}}}\left( {{{\mathbf{S}}_{\text{c}}}} \right),{\kern 1pt} {\kern 1pt} {\kern 1pt} {{\mathbf{S}}_{\text{s}}} = {{\cal R}_{{{\bm{\varphi }}_{\text{s}}}}}\left( {{{\mathbf{S}}_{\text{t}}}} \right),{\kern 1pt} {\kern 1pt} {\kern 1pt} {{\mathbf{S}}_{\text{r}}} = {{\cal D}_{{{\bm{\varphi }}_{\text{r}}}}}\left( {{{\mathbf{S}}_{\text{s}}}} \right),
\end{align}
where ${{{\bm{\varphi }}_{\text{t}}}}$, ${{{\bm{\varphi }}_{\text{s}}}}$, and ${{{\bm{\varphi }}_{\text{r}}}}$ represents the corresponding WPNN parameters at the encoder, the intermediate mapper, and  the decoder, respectively.

\section{WPNN Implementation}
In this section, we first discuss how to implement ${{{\bm{\varphi }}_{\text{t}}}}$, ${{{\bm{\varphi }}_{\text{s}}}}$, and ${{{\bm{\varphi }}_{\text{r}}}}$  using the relay and  AI-SIMs.   We then present the training strategy and loss functions designed to  train the WPNNs.

\subsection{The Architecture of the WPNNs}
According to \eqref{channel: transmitter}, the signal emitted by the transmitter-side AI-SIM is given by 
\begin{align}
{\mathbf{S}}_{\text{t}} = {\mathbf{X}}_{L}^{\text{t}}.
\end{align}
Then, the signal received at the relay, denoted by ${{\mathbf{Y}}_{\text{t}}}\in {{\mathbb C}^{{N_{\rm s}} \times \frac{{CHW}}{{2{N_{\text{t}}}}}}}$, is given by 
\begin{align}
{{\mathbf{Y}}_{\text{t}}} = {{\mathbf{G}}_{\text{t}}}{{\mathbf{S}}_{\text{t}}} + {{\mathbf{N}}_{\text{t}}},
\end{align}
where ${\left[ {{{\mathbf{N}}_{\text{t}}}} \right]_{i,j}} \sim {\cal CN}\left( {0,{\sigma ^2}} \right)$ represents the additive white Gaussian
noise.

To recover the distorted signal ${{\mathbf{S}}_{\text{t}}}$, a least squares (LS) estimator is adopted, yielding 
\begin{align}
{{{\mathbf{\hat S}}}_{\text{t}}} = {\mathbf{G}}_{\text{t}}^\dag {{\mathbf{Y}}_{\text{t}}} = {\mathbf{G}}_{\text{t}}^\dag \left( {{{\mathbf{G}}_{\text{t}}}{{\mathbf{S}}_{\text{t}}} + {{\mathbf{N}}_{\text{t}}}} \right).
\end{align} 
Subsequently, the relay  amplifies the estimated signal and forwards it to the receiver. Thus, the signal transmitted by the relay is given by 
\begin{align}
	{{\mathbf{S}}_{\text{s}}} = \tilde \pi \left( {{\mathbf{{\bf Z}}}{{{\mathbf{\hat S}}}_{\text{t}}} + {{\mathbf{1}}^T} \otimes {\mathbf{b}}_1^{\text{s}}} \right),
\end{align}
where ${\mathbf{{\bf Z}}}\in {{\mathbb C}^{{N_{\text{s}}} \times M}} $ represents the relay amplification matrix and  ${\mathbf{b}}_1^{\text{s}} \in {{\mathbb C}^{{N_{\text{s}}} \times 1}}$ denotes the bias vector implemented by injecting a controllable direct-current offset at the relay. In addition, $\tilde \pi \left(  \cdot  \right)$ represents the nonlinear function induced by the imperfect power amplifier, where the output power of  the amplifier  saturates when the input power is high. In this paper, we adopt the Rapp model for characterizing the nonlinear power amplifier behavior, which  is modelled as \cite{rapp1991effects}
\begin{align}
\tilde \pi \left( x \right) = \frac{x}{{{{\left( {1 + {{\left( {{{\left| x \right|} \mathord{\left/
										{\vphantom {{\left| x \right|} {{x_{{\text{sat}}}}}}} \right.
										\kern-\nulldelimiterspace} {{x_{{\text{sat}}}}}}} \right)}^{2p}}} \right)}^{{1 \mathord{\left/
						{\vphantom {1 {\left( {2p} \right)}}} \right.
						\kern-\nulldelimiterspace} {\left( {2p} \right)}}}}}},
\end{align} 
where $p$ and $x_{\rm sat}$ denote the circuit parameters. Following \cite{bergel2024nonlinear}, we set  $p=2$ and $x_{\rm sat}=1$, which results in a saturating amplitude response resembling the shape of a tanh function.

\textbf{\textit{Remark:}} Although the nonlinear characteristic of the relay power amplifier can be approximated by a tanh-like function, it originates from intrinsic device saturation and represents an undesired hardware impairment. In contrast, the tanh-like nonlinearity in the activation metasurface is deliberately implemented via power detection and voltage-controlled gain modulation, forming a programmable and trainable activation operator at the physical layer.

The signals then propagate through the channel ${{\mathbf{G}}_{\text{r}}}$ and the receiver-side AI-SIM, and are finally collected by the receiver antenna array. Based on \eqref{channel: receiver},  we have ${\mathbf{X}}_0^{\text{r}}={{\mathbf{G}}_{\text{r}}}{{\mathbf{S}}_{\text{s}}}$, yielding
\begin{align}
{{\mathbf{S}}_{\text{r}}} = {{\mathbf{W}}_{2L+1}^{\text{r}}}{{\mathbf{X}}_L^{\text{r}}} + {{\mathbf{N}}_{\text{r}}}  \in {{\mathbb C}^{{N_{\text{r}}} \times \frac{{CHW}}{{2{N_{\text{t}}}}}}},
\end{align}
where ${\left[ {{{\mathbf{N}}_{\text{r}}}} \right]_{i,j}} \sim {\cal CN}\left( {0,{\sigma ^2}} \right)$ represents additive white Gaussian noise and ${{\mathbf{W}}_{2L+1}^{\text{r}}}$ denotes the channel from the last layer of the receiver side AI-SIM to the receiver antennas.

Finally, the complex-valued matrix ${{\mathbf{S}}_{\text{r}}}$ is converted into its real-valued representation by concatenating the real and imaginary parts. The resulting real-valued tensor is then flattened into a one-dimensional feature vector, which is fed into an  FC layer followed by softmax activation for  image classification.  

\subsection{Training Strategy}

The task is to perform image classification at the receiver. A standard cross-entropy loss function is adopted to train the WPNNs, expressed as
\begin{align}
{{\cal L}_{{\text{loss}}}}\left( {{\mathbf{Z}},{\mathbf{b}}_1^{\text{s}},{\mathbf{\Theta }}_l^{\text{t}},{\mathbf{b}}_l^{\text{t}},{\mathbf{\Theta }}_l^{\text{r}},{\mathbf{b}}_l^{\text{r}}} \right) =  - \sum\limits_{i = 1}^C {{p_i}\log \left( {{{\hat p}_i}} \right)} ,
\end{align}
where $C$ denotes the number of classes, ${{p_i}}$ and ${{{\hat p}_i}}$ denote the true label and the predicted  probability of the $i$th class, respectively.

\section{Numerical Results}
In this section, we present numerical results to evaluate the performance of the proposed relay-assisted AI-SIM architecture for implementing WPNN.   
 We adopt the Fashion-MNIST dataset for performance evaluation, which consists of 70,000 grayscale images of size $28\times28$ from  10  categories, including 
60,000 training samples and 10,000 test samples. The SNR is defined as ${\rm{SNR = 10lo}}{{\rm{g}}_{10}}\frac{{{1}}}{{{\sigma ^2}}}$ in dB. 
The carrier frequency is set to $f_{\rm c}=2.2$~GHz, which corresponds to a wavelength $\lambda  = \frac{c}{{{f_{\text{c}}}}}$, where $c$ denotes the speed of light.
Each metasurface layer consists of uniformly spaced meta-atoms with an inter-element spacing of ${\lambda  \mathord{\left/
		{\vphantom {\lambda  2}} \right.
		\kern-\nulldelimiterspace} 2}$.  The ULA at the transmitter and receiver is placed at a distance of $10\lambda$ from the first AI-SIM layer. In addition, adjacent metasurface layers are separated by a distance of $2\lambda$.
Unless otherwise specified, the default system parameters are set as $A = {{{\lambda ^2}} \mathord{\left/
		{\vphantom {{{\lambda ^2}} 4}} \right.
		\kern-\nulldelimiterspace} 4}$,  $L=1$, $N_{\rm t}=N_{\rm s}=N_{\rm r}=14$, and $M=16$.

To quantify the performance gain achieved by the proposed scheme, we consider the following schemes:
\begin{itemize}
\item \textbf{Relay, nonlinear (Proposed):} The proposed relay-assisted SIM-enabled WPNN, where both the AI-SIM and the relay  power amplifier implement  tanh-type nonlinear operations.
\item \textbf{Relay, linear:}  A linear counterpart of the proposed scheme, where the activation metasurface is disabled and the relay power amplifier  is assumed to be linear.
\item \textbf{Relay, W/o CP:}  The same as the proposed scheme but without relay-side channel processing.  The relay directly performs forward-and-amplify protocol without LS  channel processing.
\item \textbf{W/o OTA, nonlinear:} An idealized baseline without OTA propagation, where the transmitter-side and receiver-side AI-SIM are directly connected without the relay and channels, while retaining nonlinear modules.
\item \textbf{W/o OTA, linear:} The same as ``W/o OTA, nonlinear" scheme but with passive SIM modules only, i.e., without activation-integrated metasurfaces.
\end{itemize}
\begin{figure}[!t]
	\centerline{\includegraphics[width=3.6in]{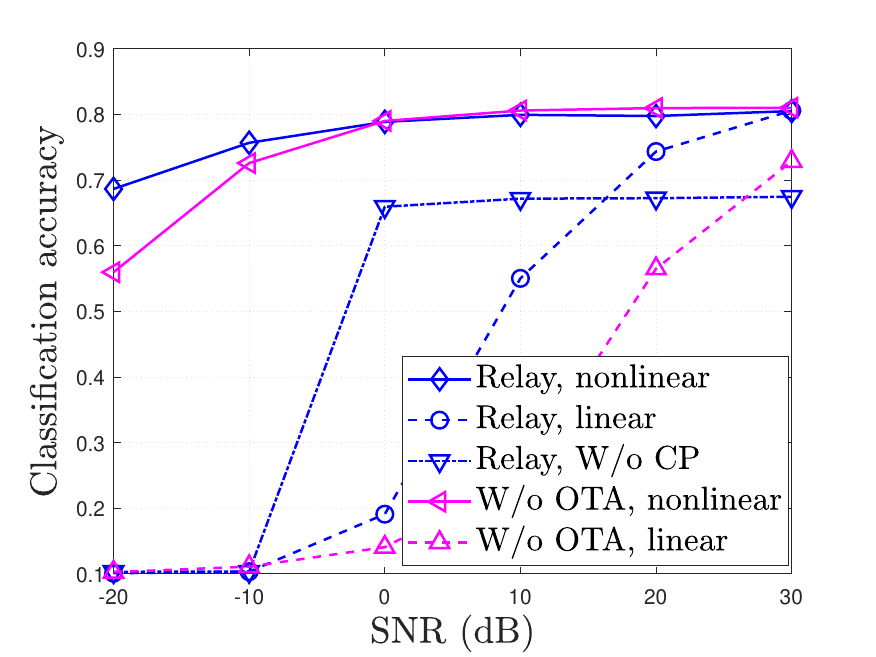}}
	\caption{SNR versus classification accuracy.} \label{fig:SNR_vs_CA}
\end{figure}
Fig.~\ref{fig:SNR_vs_CA} compares the classification accuracy of different schemes versus SNR. The proposed ``Relay, nonlinear'' scheme exhibits strong robustness in the low-SNR regime and achieves  a classification accuracy of 0.6867  even at ${\rm SNR}=-20$~dB, validating that the relay-assisted AI-SIM  architecture can effectively perform WPNN inference over noisy wireless links. 
By contrast, removing physical-layer activation in ``Relay, linear'' scheme leads to severe performance degradation at low SNRs, where the classification accuracy drops to 0.1011 at ${\rm SNR}=-20$~dB, indicating the necessity of nonlinear activation. Moreover, without LS-based relay processing, the ``Relay, W/o CP'' scheme yields very low accuracy when 
${\rm SNR}\le 0$~dB, highlighting the importance of relay-side signal processing against the channel variation. Finally, compared with the idealized ``W/o OTA, nonlinear'' baseline, the proposed scheme achieves comparable overall performance and even slightly outperforms it in the very low-SNR regime. This performance gain can be attributed to the implicit robustness induced by OTA propagation and relay-assisted processing.

\begin{figure}[!t]
	\centerline{\includegraphics[width=3.6in]{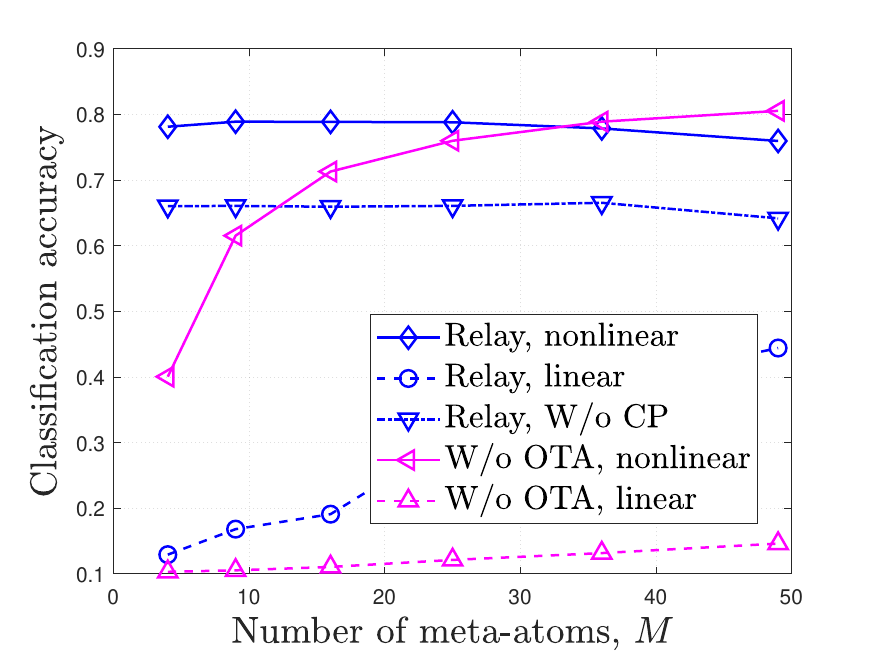}}
	\caption{Number of meta-atoms versus classification accuracy.} \label{fig:RISelement_vs_AC}
\end{figure}

Fig.~\ref{fig:RISelement_vs_AC} shows the classification accuracy versus the number of meta-atoms $M$ under ${\rm SNR}=0$~dB. The proposed ``Relay, nonlinear'' scheme achieves the best performance when $M\le36$ and consistently outperforms the other benchmark schemes in this regime. In particular, when $M=4$, the proposed scheme attains a classification accuracy of $0.7815$. This improvement can be attributed to the relay-assisted architecture effectively forming an additional FC layer, thereby increasing the depth of the WPNN.
As $M$ increases, the performance of the proposed scheme remains nearly unchanged. This behavior suggests that the overall WPNN performance is primarily constrained by the antenna dimensions, since the relay-induced FC layer dominates the effective network capacity. This observation is further supported by the ``W/o OTA, nonlinear'' scheme, where the classification accuracy improves with increasing $M$, indicating that in the absence of OTA constraints, enlarging the SIM dimension directly enhances the representation capability.

\begin{figure}[!t]
	\centerline{\includegraphics[width=3.6in]{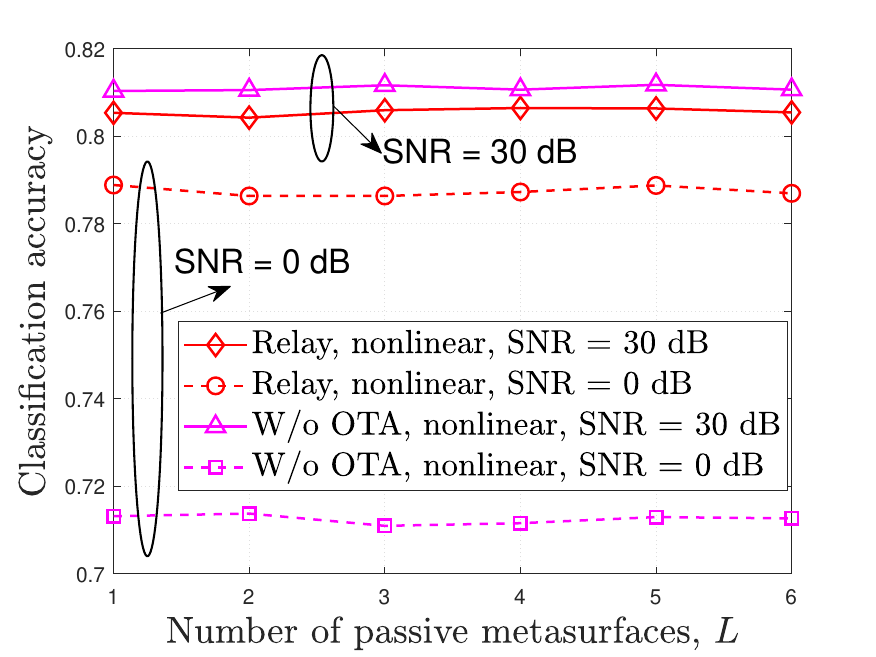}}
	\caption{Number of passive metasurfaces versus classification accuracy.} \label{fig:NumberofRISs_vs_AC}
\end{figure}

Fig.~\ref{fig:NumberofRISs_vs_AC}  illustrates the classification accuracy versus the number of passive metasurfaces $L$ under different SNR conditions. Increasing $L$, which effectively increases the depth of the WPNNs, leads to nearly unchanged classification accuracy across all considered SNRs.  It is worth noting that even when $L=1$, the effective network already consists of four FC layers, including two layers formed by the AI-SIM modules, one layer induced by the relay, and a final FC layer at the receiver for classification, implying that the network depth is sufficient for accurate inference on the considered task.  In addition, it can be observed that when $\mathrm{SNR}=30~\mathrm{dB}$, the proposed scheme  closely approaches the ``W/o OTA, nonlinear'' scheme. In contrast, when $\mathrm{SNR}=0~\mathrm{dB}$, the proposed scheme achieves significantly higher classification accuracy, mainly because the relay provides additional degrees of freedom beyond the unit-modulus, phase-only control of the SIM, enabling more effective noise suppression under low-SNR conditions.

\begin{figure}[!t]
	\centerline{\includegraphics[width=3.6in]{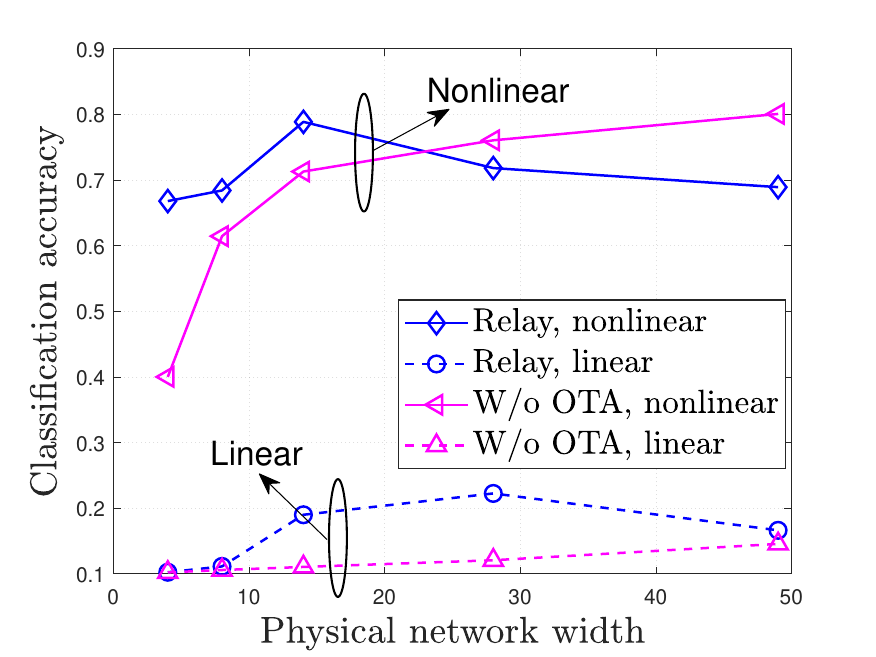}}
	\caption{Physical network width versus classification accuracy.} \label{fig:Width_vs_CA}
\end{figure}

Fig.~\ref{fig:Width_vs_CA} shows the classification accuracy versus the physical network width under $\mathrm{SNR}=0~\mathrm{dB}$, where the widths of the AI-SIM, relay, and transceiver are set to be identical. As the width increases, the nonlinear schemes exhibit a clear advantage over their linear counterparts, confirming the essential role of physical-layer nonlinearity for reliable inference. In particular, the proposed ``Relay, nonlinear'' scheme improves markedly when increasing the width from small values and achieves its best performance at a moderate width of 16, while a further increase in width leads to a slight performance degradation. This non-monotonic behavior indicates a tradeoff between dimensionality gain and noise accumulation in relay-assisted OTA WPNNs. 
By contrast, the idealized ``W/o OTA, nonlinear'' baseline improves steadily with the width and attains the highest accuracy at large widths as expected.

\section{Conclusion}

This paper proposed a relay-assisted WPNN architecture enabled by AI-SIMs. By jointly exploiting  trainable metasurface phase shifts, relay amplification coefficients, and hardware-implemented nonlinear activations, the proposed framework enabled end-to-end inference directly in the physical layer. Numerical results showed that the proposed architecture significantly improves classification accuracy over purely linear physical implementations, particularly in low-SNR regimes. Moreover, the effects of network width and depth were examined, revealing inherent tradeoffs between dimensionality gain and noise accumulation in relay-assisted OTA WPNNs.

\bibliographystyle{IEEEtran}
\bibliography{PNNRISRelay}

@Article{bergel2024nonlinear,
  Title                    = {Non-Linear Relay Optimization Using Deep-Learning Tools},
  Author                   = {Bergel, Itsik},
  Journal                  = {IEEE Trans. Wireless Commun.},
  Year                     = {2024},

  Month                    = {Dec.},
  Number                   = {12},
  Pages                    = {19289-19301},
  Volume                   = {23}
}

@Article{bian2025overtheair,
  Title                    = {Over-the-Air Inference Through Analog Computation Over Multi-hop {MIMO} Networks},
  Author                   = {Bian, Chenghong and Hua, Meng and Gündüz, Deniz},
  Journal                  = {IEEE Wireless Commun. Lett.},
  Year                     = {2025, early access, doi: 10.1109/LWC.2025.3602209}
}

@InProceedings{binyamini2025estimating,
  Title                    = {Estimating and Optimizing of Deep Relay Networks},
  Author                   = {Binyamini, Ido and Bergel, Itsik},
  Booktitle                = {Proc. IEEE 26th Int. Workshop Signal Process. Artif. Intell. Wireless Commun. (SPAWC)},
  Year                     = {2025},
  Pages                    = {1--5}
}

@ARTICLE{hua2025aircnn,
  author={Hua, Meng and Wu, Haotian and Gündüz, Deniz},
  journal={IEEE Wireless Communications Letters}, 
  title={{CNNs} in the Air via Reconfigurable Intelligent Surfaces}, 
  year={2026},
  month={Mar.},
  volume={15},
  number={},
  pages={2124-2128},
}

@article{hua2026WPNN,
  title={Wireless physical neural networks {(WPNNs)}: Opportunities and challenges},
  author={Hua, Meng and Bergel, Itsik and Girici, Tolga and Di Renzo, Marco and Gunduz, Deniz},

  year={2026},
    Publisher                = {[Online]. Available:},
  Url                      = {https://arxiv.org/abs/2602.14094.}
}

@article{girici2026realization,
  title={Realization of a Fully Connected Neural Layer Over-the-Air through Multi-hop Amplify-and-Forward Relays},
  author={Girici, Tolga and Hua, Meng and G{\"u}nd{\"u}z, Deniz},
 
  year={2026},
     Publisher                = {[Online]. Available:},
  Url                      = {https://arxiv.org/abs/2603.20489.}
}

@Article{iten2020discovering,
  Title                    = {Discovering Physical Concepts with Neural Networks},
  Author                   = {Iten, Raban and Metger, Tony and Wilming, Henrik and del Rio, L\'{\i}dia and Renner, Renato},
  Journal                  = {Phys. Rev. Lett.},
  Year                     = {2020},

  Month                    = {Jan},
  Pages                    = {010508},
  Volume                   = {124},

  Issue                    = {1},
  Numpages                 = {6}
}

@Article{ning2025multilayer,
  Title                    = {Multilayer nonlinear diffraction neural networks with programmable and fast {ReLU} activation function},
  Author                   = {Ning, Yu Ming and Ma, Qian and Xiao, Qiang and Gao, Xin Xin and Wu, Qian Wen and Gu, Ze and Li, Rui Si and Chen, Long and You, Jian Wei and Cui, Tie Jun},
  Journal                  = {Nat. Commun.},
  Year                     = {2025},
  Number                   = {1},
  Pages                    = {10332},
  Volume                   = {16}
}

@Article{rapp1991effects,
  Title                    = {Effects of HPA-nonlinearity on a {4-DPSK/OFDM}-signal for a digital sound broadcasting signal},
  Author                   = {Rapp, Christoph},
  Journal                  = { Proc. 2nd Eur. Conf. Satell.Commun.},
  Year                     = {1991},
  Pages                    = {179--184},
  Volume                   = {332}
}

@Article{Garcia2023irNN,
  Title                    = {{AirNN}: Over-the-Air Computation for Neural Networks via Reconfigurable Intelligent Surfaces},
  Author                   = {Garcia Sanchez and others},
  Journal                  = {IEEE/ACM Tran. Netw.},
  Year                     = {2023},

  Month                    = {Dec.},
  Number                   = {6},
  Pages                    = {2470-2482},
  Volume                   = {31}
}

@Article{stylianopoulos2025over,
  Title                    = {Over-the-Air Edge Inference via End-to-End Metasurfaces-Integrated Artificial Neural Networks},
  Author                   = {Stylianopoulos, Kyriakos and Di Lorenzo, Paolo and Alexandropoulos, George C},
  Year                     = {2025},

  Publisher                = {[Online]. Available:},
  Url                      = {https://arxiv.org/abs/2504.00233.}
}

@Article{stylianopoulos2026over,
  Title                    = {Over-The-Air Extreme Learning Machines with {XL} Reception via Nonlinear Cascaded Metasurfaces},
  Author                   = {Stylianopoulos, Kyriakos and Fabiani, Mattia and Torcolacci, Giulia and Dardari, Davide and Alexandropoulos, George C},
  Year                     = {2026},

  Publisher                = {[Online]. Available:},
  Url                      = {https://arxiv.org/abs/2601.17749.}
}

@Article{wang2022distributed,
  Title                    = {Distributed Learning for {MIMO} Relay Networks},
  Author                   = {Wang, Rui and Jiang, Yi and Zhang, Wei},
  Journal                  = {IEEE J. Sel. Topics Signal Process.},
  Year                     = {2022},

  Month                    = {Apr.},
  Number                   = {3},
  Pages                    = {343-357},
  Volume                   = {16}
}

@Article{wright2022deep,
  Title                    = {Deep physical neural networks trained with backpropagation},
  Author                   = {Wright, Logan G. and Onodera, Tatsuhiro and Stein, Martin M. and Wang, Tianyu and Schachter, Darren T. and Hu, Zoey and McMahon, Peter L.},
  Journal                  = {Nature},
  Year                     = {2022},
  Number                   = {7894},
  Pages                    = {549--555},
  Volume                   = {601}
}
\end{document}